\documentclass[twocolumn]{aastex63}

\usepackage{amsmath}
\usepackage{amssymb}
\usepackage{rotating}

\usepackage[T1]{fontenc}
\usepackage[utf8]{inputenc}

\newcommand{\ra}[4]{${#1}^{\rm h}{#2}^{\rm m}{#3}\fs{#4}$}
\newcommand{\dec}[4]{${#1}\arcdeg{#2}\arcmin{#3}\farcs{#4}$}
\newcommand{\rashort}[3]{${#1}^{\rm h}{#2}^{\rm m}{#3}^{\rm s}$}
\newcommand{\decshort}[3]{${#1}\arcdeg{#2}\arcmin{#3}\arcsec$}

\newcommand\tE{t_{\rm E}}

\newcommand\murel{\mu_{\rm rel}}

\newcommand\pirel{\pi_{\rm rel}}
\newcommand\thetaE{\theta_{\rm E}}

\received{}
\revised{}
\accepted{}

\shorttitle{Limits on planetary-mass primordial black holes}
\shortauthors{P. Mr\'oz et al.}

\begin{document}

\title{Limits on planetary-mass primordial black holes from the OGLE high-cadence survey of the Magellanic Clouds}

\correspondingauthor{Przemek Mr\'oz}
\email{pmroz@astrouw.edu.pl}

\author[0000-0001-7016-1692]{Przemek Mr\'oz}
\affil{Astronomical Observatory, University of Warsaw, Al. Ujazdowskie 4, 00-478 Warszawa, Poland}

\author[0000-0001-5207-5619]{Andrzej Udalski}
\affil{Astronomical Observatory, University of Warsaw, Al. Ujazdowskie 4, 00-478 Warszawa, Poland}

\author[0000-0002-0548-8995]{Micha\l{} K. Szyma\'nski}
\affil{Astronomical Observatory, University of Warsaw, Al. Ujazdowskie 4, 00-478 Warszawa, Poland}

\author[0000-0002-7777-0842]{Igor Soszy\'nski}
\affil{Astronomical Observatory, University of Warsaw, Al. Ujazdowskie 4, 00-478 Warszawa, Poland}

\author[0000-0002-2339-5899]{Pawe\l{} Pietrukowicz}
\affil{Astronomical Observatory, University of Warsaw, Al. Ujazdowskie 4, 00-478 Warszawa, Poland}

\author[0000-0003-4084-880X]{Szymon Koz\l{}owski}
\affil{Astronomical Observatory, University of Warsaw, Al. Ujazdowskie 4, 00-478 Warszawa, Poland}

\author[0000-0002-9245-6368]{Rados\l{}aw Poleski}
\affil{Astronomical Observatory, University of Warsaw, Al. Ujazdowskie 4, 00-478 Warszawa, Poland}

\author[0000-0002-2335-1730]{Jan Skowron}
\affil{Astronomical Observatory, University of Warsaw, Al. Ujazdowskie 4, 00-478 Warszawa, Poland}

\author[0000-0001-6364-408X]{Krzysztof Ulaczyk}
\affil{Department of Physics, University of Warwick, Coventry CV4 7 AL, UK}
\affil{Astronomical Observatory, University of Warsaw, Al. Ujazdowskie 4, 00-478 Warszawa, Poland}

\author[0000-0002-1650-1518]{Mariusz Gromadzki}
\affil{Astronomical Observatory, University of Warsaw, Al. Ujazdowskie 4, 00-478 Warszawa, Poland}

\author[0000-0002-9326-9329]{Krzysztof Rybicki}
\affil{Department of Particle Physics and Astrophysics, Weizmann Institute of Science, Rehovot 76100, Israel}
\affil{Astronomical Observatory, University of Warsaw, Al. Ujazdowskie 4, 00-478 Warszawa, Poland}

\author[0000-0002-6212-7221]{Patryk Iwanek}
\affil{Astronomical Observatory, University of Warsaw, Al. Ujazdowskie 4, 00-478 Warszawa, Poland}

\author[0000-0002-3051-274X]{Marcin Wrona}
\affil{Department of Astrophysics and Planetary Sciences, Villanova University, 800 Lancaster Ave., Villanova, PA 19085, USA}
\affil{Astronomical Observatory, University of Warsaw, Al. Ujazdowskie 4, 00-478 Warszawa, Poland}

\author{Mateusz J. Mr\'oz}
\affil{Astronomical Observatory, University of Warsaw, Al. Ujazdowskie 4, 00-478 Warszawa, Poland}

\begin{abstract}
Observations of the Galactic bulge revealed an excess of short-timescale gravitational microlensing events that are generally attributed to a large population of free-floating or wide-orbit exoplanets. However, in recent years, some authors suggested that planetary-mass primordial black holes (PBHs) comprising a substantial fraction (1\%--10\%) of the dark matter in the Milky Way may be responsible for these events. If that was the case, a large number of short-timescale microlensing events should also be seen toward the Magellanic Clouds. Here, we report the results of a high-cadence survey of the Magellanic Clouds carried out from 2022 October through 2024 May as part of the Optical Gravitational Lensing Experiment. We observed almost 35~million source stars located in the central regions of the Large and Small Magellanic Clouds and found only one long-timescale microlensing event candidate. No short-timescale events were detected despite high sensitivity to such events. That allows us to infer the strongest available limits on the frequency of planetary-mass PBHs in dark matter. We find that PBHs and other compact objects with masses from $1.4 \times 10^{-8}\,M_{\odot}$ (half of the Moon mass) to $0.013\,M_{\odot}$ (planet/brown dwarf boundary) may comprise at most 1\% of dark matter. That rules out the PBH origin hypothesis for the short-timescale events detected toward the Galactic bulge and indicates they are caused by the population of free-floating or wide-orbit planets.
\end{abstract}


\section{Introduction} \label{sec:intro}

High-cadence observations of the Galactic bulge carried out by gravitational microlensing surveys in the past two decades have revealed an excess of short-timescale microlensing events compared to the expectations from brown dwarf and stellar populations \citep{mroz2017,gould2022,sumi2023}. These events are characterized by short Einstein timescales \mbox{$\tE \lesssim 0.5$\,day} and small angular Einstein radii $\thetaE \lesssim 10\,\mu\mathrm{as}$. They are commonly attributed to a population of free-floating (that is, gravitationally unbound) or wide-orbit exoplanets (hereafter, FFPs) in the Galactic disk and bulge. Because both the angular Einstein radius and the Einstein timescale are proportional to the square root of the lens mass $M$ as
\begin{align}
\thetaE &\approx 10\,\mu\mathrm{as}\,\left(\frac{M}{10\,M_{\oplus}}\frac{\pirel}{0.05\,\mathrm{mas}}\right)^{1/2}, \\
\tE &\approx 0.5\,\mathrm{day}\,\left(\frac{M}{10\,M_{\oplus}}\frac{\pirel}{0.05\,\mathrm{mas}}\right)^{1/2}\left(\frac{\murel}{7\,\mathrm{mas\,yr}^{-1}}\right)^{-1},
\end{align}
masses of FFPs are estimated to be a few $M_{\oplus}$. Here, $\pirel$ is the relative lens--source parallax, and $\murel$ is the relative lens--source proper motion.

All three major studies on FFPs \citep{mroz2017,gould2022,sumi2023} agree that FFPs seem to be frequent in the Milky Way; there are several FFPs per every star. The distributions of timescales and Einstein radii of detected short-duration microlensing events are consistent with a power-law mass function of FFPs: $dN/d\log M \approx 0.5 (M/38\,M_{\oplus})^{-p}$ per star with $p\approx 1$ for $M \gtrsim 1\,M_{\oplus}$ \citep{gould2022,sumi2023}. That corresponds to the total mass in the form of FFPs of $80\!-\!200\,M_{\oplus}$ per star. Assuming a mean stellar mass of $0.5\,M_{\odot}$, the total mass of stars in the Milky Way of $5\times 10^{10}\,M_{\odot}$, and the total dark matter halo mass of the Milky Way of $9.5 \times 10^{11}\,M_{\odot}$ \citep{cautun2020}, the ratio of the total mass of FFPs to the dark matter halo mass is $f_{\rm FFP} \approx (2\!-\!6)\times 10^{-4}$.

\begin{figure*}[htb]
\centering
\includegraphics[width=.7\textwidth]{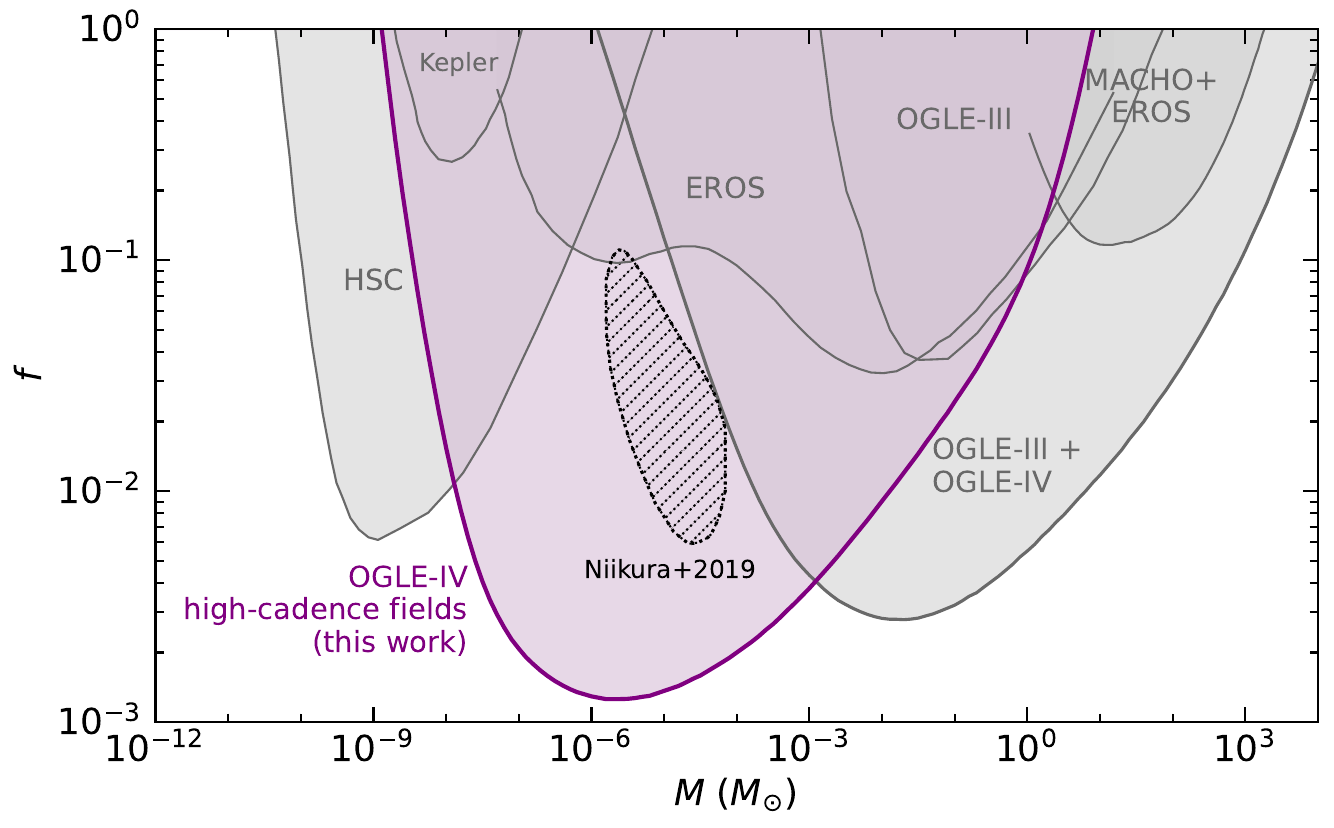}
\caption{The 95\% upper limits on the fraction of dark matter in the form of PBHs and other compact objects. The hatched area marks the region of the PBH parameter space obtained by \citet{niikura2019b}, under assumption that short-timescale microlensing events detected by \citet{mroz2017} in the OGLE-IV Galactic bulge high-cadence survey are due to planetary-mass PBHs. The shaded purple region corresponds to the 95\% confidence limit on $f$ found in this work. Other limits are adopted from \citet{tisserand2007} (EROS), \citet{wyrzyk1} (OGLE-III), \citet{griest2013} (Kepler), \citet{niikura2019} (HSC), \citet{moniez2022} (MACHO+EROS), and \citet{mroz2024b} (OGLE-III+OGLE-IV).}
\label{fig:bounds}
\end{figure*}

Soon after the discovery of six short-timescale microlensing events by \citet{mroz2017} in the data from the high-cadence Galactic bulge survey by the fourth phase of the Optical Gravitational Lensing Experiment (OGLE), the paper by \citet{niikura2019b} raised speculations about their origin. These authors put forward a claim that the short-timescale events discovered by OGLE were caused by planetary-mass primordial black holes (PBHs), not free-floating or wide-orbit planets, as argued in the original discovery paper. \citet{niikura2019b} found that if the lenses were indeed PBHs and if their distribution in the Milky Way followed the Navarro--Frenk--White model \citep{navarro1997,klypin2002}, then PBHs of a few $M_{\oplus}$ could comprise $1\%\!-\!10\%$ of the dark matter in the Milky Way (see the hatched area in Figure~\ref{fig:bounds}).

These claims inspired many theoretical follow-up studies, which attempted to explain the origin of planetary-mass PBHs allegedly detected in the OGLE data. For example, a purported population of planetary-mass PBHs was claimed to be consistent with different models of inflation \citep[e.g.,][]{tada2019, fu2019, motohashi2020, teimoori2021b, teimoori2021, solbi2021, heydari2022, zhang2022, yi2023, fu2023, fu2023b, flores2023, ashrafzadeh2024, heydari2024, heydari2024b, yi2024, cai2024, yang2024}. Several authors \citep[e.g.,][]{domenech2022, zhang2022, yi2023, franciolini2023, inomata2024, domenech2024, yang2024, yi2024} proposed that planetary-mass PBHs may be linked to the detection of the stochastic gravitational wave background \citep{arzoumanian2020,agazie2023,agazie2023b,epta2023b,epta2023,reardon2023,zic2023,xu2023}, as they both can have a common origin in the primordial curvature perturbations.

Planetary-mass PBHs can also be produced during first-order electroweak phase transitions in the early Universe \citep[e.g.,][]{hashino2022,hasino2023,kawana2023,gouttenoire2024,goncalves2024}, vacuum transitions \citep{kawana2023}, QCD phase transitions \citep{lu2023}, or by the collapse of domain walls \citep[e.g.,][]{ge2024}. In particular, \citet{carr2021b} proposed that planetary-mass PBHs may have been formed as a result of the sudden drop in the pressure of relativistic matter caused by the $W^{\pm}/Z^0$ boson decoupling in the early Universe.

Short-timescale microlensing events detected by OGLE were also speculated to be caused by axion stars \citep{sugiyama2023} or dark soliton stars \citep{del_grosso_2024}. The idea put forward by \citet{niikura2019b} also led to claims that the proposed Planet~9 \citep{batygin2016,batygin2019} may actually be a PBH \citep{scholtz2020}.

We expect that planets (either free-floating or bound to their host stars) should follow the distribution of stars in the Milky Way, whereas PBHs would track the distribution of dark matter. Distinguishing between the free-floating/wide-orbit planet and the PBH hypotheses for Galactic bulge microlensing events would require precise measurements of the distances to lensing objects.\footnote{The populations of FFPs and PBHs can, in theory, also be separated in a statistical sense based on $\tE$ or $\thetaE$ measurements for a large sample of events \citep[e.g.,][]{derocco2024}.} That would, in turn, require measurements of microlensing parallaxes of short-timescale events, which are, in practice, nearly impossible with the current technology. (Such measurements may be feasible in the future with simultaneous observations from two separated observatories; e.g., \citealt{zhu_gould_2016,bachelet2019,ban2020,gould2021,yan2022}.)

An obvious solution to this problem is to search for short-timescale microlensing events outside the Galactic bulge and disk, where the signal from the Milky Way stellar populations is expected to be negligible. Such experiments have been conducted toward the Magellanic Clouds by the EROS survey \citep{renault1997,renault1998,tisserand2007} and toward the Andromeda Galaxy by the Subaru Hyper Suprime-Cam (HSC) survey \citep{niikura2019} and the Canada--France--Hawaii Telescope \citep{gu2024}, but they were not sensitive enough to the signal found by \citet{niikura2019b}, as shown in Figure~\ref{fig:bounds}. Similarly, the combined \mbox{OGLE-III} and \mbox{OGLE-IV} observations of the Large Magellanic Cloud (LMC; \citealt{mroz2024a,mroz2024b}) were not sensitive enough to short-timescale events expected from planetary-mass PBHs (Figure~\ref{fig:bounds}).

We, therefore, decided to conduct a high-cadence survey of the Magellanic Clouds to verify this speculative (but potentially highly rewarding) idea that a substantial fraction of dark matter is made of planetary-mass PBHs. This article presents the results of our survey.

\section{Data}

The data analyzed in this study were collected as part of the OGLE-IV survey \citep{udalski2015}. The survey uses the dedicated 1.3~m Warsaw Telescope located at Las Campanas Observatory, Chile. (The observatory is operated by the Carnegie Institution for Science.) The telescope is equipped with a mosaic CCD camera comprising thirty-two $2048 \times 4102$ detectors. The pixel scale is 0.26\,arcsec\,pixel$^{-1}$, providing a total field of view of 1.4\,deg$^2$. \citet{udalski2015} presented a detailed description of this instrument.

\begin{figure*}
\centering
\includegraphics[width=\textwidth]{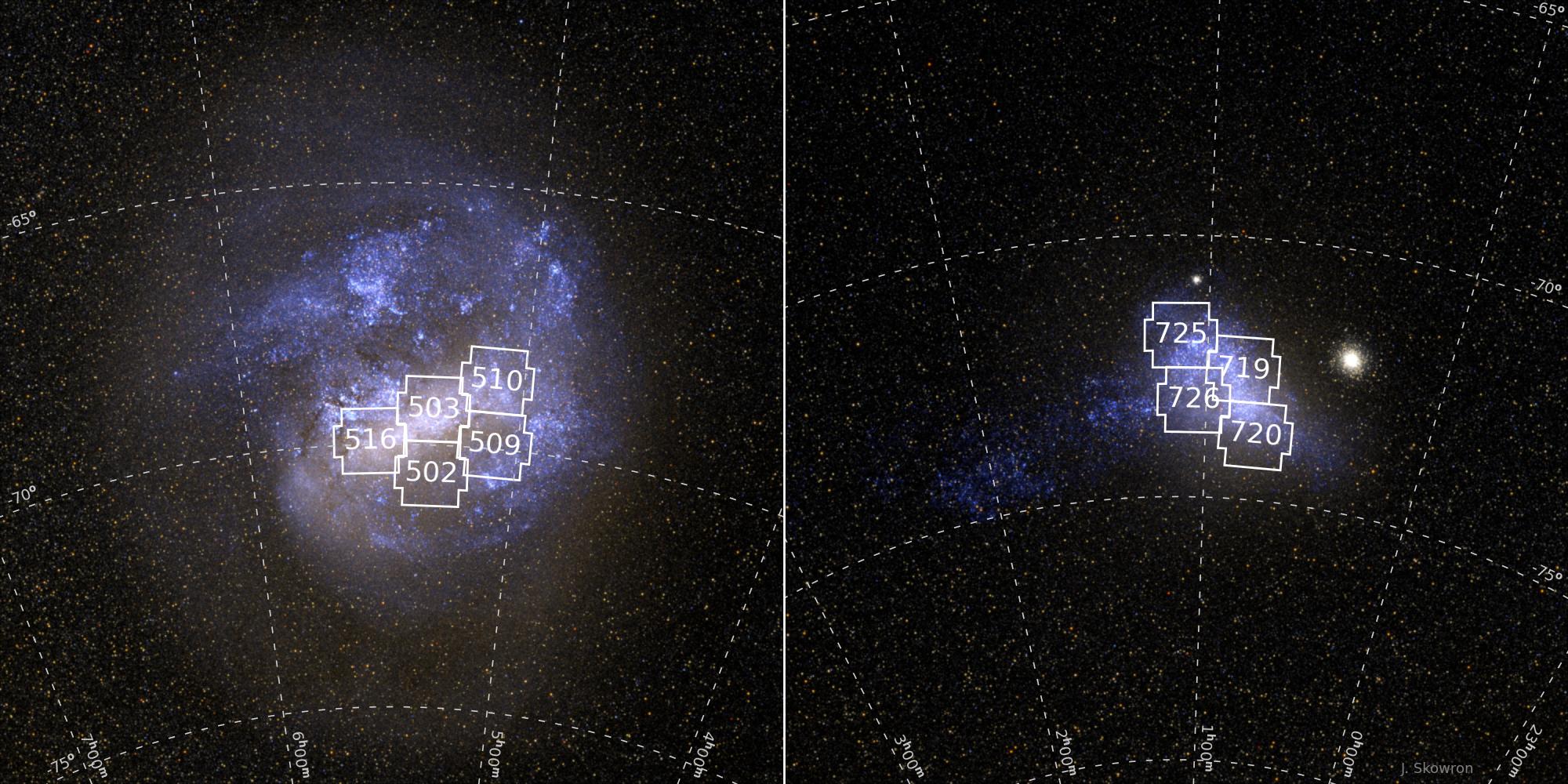}
\caption{Fields of the OGLE High-cadence Magellanic Clouds Survey; left panel---LMC; right panel---SMC. The background images of the LMC and SMC were generated with \textsc{bsrender} written by Kevin Loch, using the ESA/Gaia database.}
\label{fig:fields}
\end{figure*}

The Magellanic Clouds have been a target of the OGLE survey since its second phase \citep{udalski1997,udalski2003,udalski2015}. However, they were observed at a moderate cadence, typically 1--5 days. When OGLE-IV operations were resumed after the COVID-19 pandemic in 2022 August, we decided to observe some Magellanic Clouds fields at high cadence. We selected four fields in the Small Magellanic Cloud (SMC) and five in the LMC, which contained the largest number of stars. They cover an area of approximately 12.4\,deg$^2$ and contain almost 35~million source stars brighter than $I=22$. Basic information about these fields (equatorial coordinates, total number of epochs in 2022/2023 and 2023/2024 observing seasons, and the number of microlensing source stars) is presented in Table~\ref{tab:fields}. The location of the fields in the sky is shown in Figure~\ref{fig:fields}.

The four SMC fields were observed up to 21 times per night (with a median of 9 times per night) with a median cadence of 16\,minutes. The median cadence in the LMC was slightly longer (20 minutes). The LMC fields were observed up to 23 times per night (with a median of 12 times per night). High-cadence survey observations of the SMC were conducted from 2022 October 9 to 2022 December 11 and from 2023 August 11 to 2024 January 29 (in total, 236 days). The LMC was observed at high cadence from 2022 October 4 to 2023 April 30 and from 2023 September 15 to 2024 April 27 (in total, 435 days).

All high-cadence data were collected in the $I$ band filter, whose transmission curve closely resembles that of the standard Cousins photometric system. In addition, a small number of $V$ band images were also collected for characterization of detected objects. Exposure times ranged from 150 to 170\,s, providing the limiting magnitude $I \approx 21.7$. We used the difference image analysis (DIA) technique \citep{tomaney1996,alard1998,wozniak2000} to extract the photometric time-series data. We refer the readers to \citet{udalski2015} for details of the photometric pipeline.

\begin{deluxetable}{lcccc}
\tablecaption{Fields of the OGLE High-cadence Magellanic Clouds Survey\label{tab:fields}}
\tablehead{
\colhead{Field} & \colhead{R.A.} & \colhead{Decl.} & \colhead{$N_{\rm epochs}$} & \colhead{$N_{\rm s}\ (\times 10^6)$}
}
\startdata
LMC502 & \rashort{05}{19}{00} & \decshort{-70}{32}{20} & 4876 & 3.26 \\
LMC503 & \rashort{05}{19}{00} & \decshort{-69}{18}{30} & 4880 & 8.21 \\
LMC509 & \rashort{05}{05}{07} & \decshort{-69}{55}{25} & 4868 & 3.77 \\
LMC510 & \rashort{05}{05}{52} & \decshort{-68}{41}{35} & 4868 & 5.99 \\
LMC516 & \rashort{05}{32}{52} & \decshort{-69}{55}{25} & 4872 & 6.36 \\
SMC719 & \rashort{00}{50}{53} & \decshort{-72}{31}{09} & 1684 & 2.44 \\
SMC720 & \rashort{00}{46}{15} & \decshort{-73}{45}{18} & 1673 & 2.00 \\
SMC725 & \rashort{01}{06}{48} & \decshort{-71}{45}{22} & 1677 & 0.79 \\
SMC726 & \rashort{01}{03}{18} & \decshort{-73}{08}{17} & 1673 & 2.07 \\
\enddata
\tablecomments{Equatorial coordinates are given for the epoch J2000.0. $N_{\rm epochs}$ is the total number of epochs in 2022/2023 and 2023/2024 observing seasons. $N_{\rm s}$ is the number of microlensing source stars (in millions) brighter than $I=22$.}
\end{deluxetable}

\section{Search for Events}
\label{sec:search}

The methods that we employed for searching for microlensing events in the time-series data are necessarily similar to those used in our previous works \citep{mroz2017,mroz2024a}. One important difference arises from the fact that we expect that some light curves of microlensing events due to planetary-mass objects may be affected by finite-source effects. The selection of events is carried out in three steps. First, we search for objects exhibiting any brightening with respect to the flat light curve. Then, we remove obvious nonmicrolensing light curves and, finally, check whether a  microlensing model can fit the light curve well. The selection criteria are devised to maximize the detection efficiency (see Section~\ref{sec:efficiency}) while keeping the contamination from nonmicrolensing light curves as small as possible. All selection cuts are summarized in Table~\ref{tab:cuts}.

\begin{table*}
\centering
\caption{Event Selection Criteria}
\label{tab:cuts}
\begin{tabular}{llr}
\hline \hline
& \multicolumn{1}{c}{Criteria} & \multicolumn{1}{c}{Number} \\
\hline
\textbf{Cut 0.} & \textbf{All stars in the databases} & 17,579,931 \\
\hline
\textbf{Cut 1.} & \textbf{Stars with at least one significant brightening in the light curve} \\
& At least five consecutive data points $3\sigma$ above the baseline flux ($n_{\rm bump} \geq 5$) & \\
& Object detected on at least three subtracted images ($n_{\rm DIA} \geq 3$) &\\
& The total ``significance'' of the bump ($\chi_{3+}=\sum_i(F_i-F_{\rm base})/\sigma_i \geq 40$) &\\
& Amplitude of the bump at least 0.05 mag ($\Delta m \geq 0.05$ mag) & \\
& No significant variability outside the window ($\chi^2_{\rm out}/\mathrm{d.o.f.} \leq 2$) & 2538 \\
\hline
\textbf{Cut 2.} & \textbf{Removing false positives} & \\
& Stars with multiple bumps in the data & \\
& ``Blue bumpers'' ($(V-I)_0 \leq 0.5$, $I_0 \leq 19.5$ in the LMC, $I_0 \leq 20.0$ in the SMC) & \\
& Photometry artifacts & 308 \\
\hline
\textbf{Cut 3.} & \textbf{Microlensing model describes the light curve well} & \\
& Fit converged & \\
& $\chi^2/\mathrm{d.o.f.} \leq 2$ (all data points) & \\
& $\chi^2_{\tE}/\mathrm{d.o.f.} \leq 2$ (for $|t - t_0| < \tE$) & \\
& $\chi^2_{2\tE}/\mathrm{d.o.f.} \leq 2$ (for $|t - t_0| < 2\tE$) & \\
& $\chi^2_{\rm bump}/\mathrm{d.o.f.} \leq 2$ (for data points within the bump) & \\
& Impact parameter is smaller than 1 or $2\rho$ ($u_0 \leq \max\{1,2\rho\}$) & \\ 
& Uncertainty on the Einstein timescale ($\sigma(\tE) / \tE \leq 1$) & \\
& Source star brighter than 22 mag ($I_{\rm s} \leq 22$) & 2\\
\hline
\end{tabular}
\end{table*}

\subsection{Cut 1}

We searched for microlensing events using only the data collected between 2022 August 9 and 2024 May 20. The error bars reported by the photometric pipeline were rescaled following the procedure described by \citet{skowron2016} and \citet{mroz2024a}, and we converted magnitudes into flux units. Then, for each light curve containing at least 30 data points, we placed a window lasting 120 days and calculated the mean flux $F_{\rm base}$ and its standard deviation $\sigma_{\rm base}$ using data points outside the window (after removing $4\sigma$ outliers). We searched for at least $n_{\rm bump} = 5$ consecutive data points within the window, whose flux was greater than $F_{\rm base} + 3\sigma_{\rm base}$. We called such data points a ``bump.'' In case such a bump was identified in the light curve, we calculated the quantity $\chi_{3+}=\sum_i (F_i - F_{\rm base})/\sigma_i$, where $(F_i,\sigma_i)$ is the flux and its uncertainty of the $i$th data point within the bump. Subsequently, the window was shifted by 40 days, and the whole procedure was repeated until the end of the light curve was reached. From all iterations of the algorithm, we selected a bump with the largest value of $\chi_{3+}$ and required $\chi_{3+} \geq 40$.

In addition, we required that during at least \mbox{$n_{\rm DIA} = 3$} epochs the magnified source was detected by the DIA pipeline on subtracted images (see, \citealt{mroz2024a}), and the amplitude of the bump should be \mbox{$\Delta m \geq 0.05$\,mag}. The latter cut was devised to remove contamination from OGLE small-amplitude red giants \citep{wray2004}, whose low-amplitude pulsations were picked up by our algorithm as candidate microlensing events. We also required that no significant variability was detected outside the magnified part of the light curve. To quantify that, we calculated $\chi^2_{\rm out}=\sum_i(F_i - F_{\rm base})^2/\sigma_i^2$, where the summation was performed over all data points outside the window and required $\chi^2_{\rm out}/\mathrm{d.o.f.} \leq 2$. In this step, we reduced the number of analyzed light curves from over 17.6 million to 2538.

\subsection{Cut 2}

In the second step, we identified false positives and removed them from the sample. They could be classified into three main (nonexclusive) classes: (1) stars with multiple bumps (brightenings) in their light curves, (2) stars located in the ``blue-bumper'' region in the color--magnitude diagram (CMD), and (3) photometry artifacts. 

First, we required that the full OGLE-III (if available) and OGLE-IV light curve contains only one bump. To verify that, we algorithmically removed magnified data points from the main brightening and ran the event-finding algorithm described above using the remaining data. If the algorithm detected another bump with $\chi_{3+} \geq 30$ and $n_{\rm bump} \geq 5$ (note these conditions are more lax than those imposed on the primary bump), we discarded such a light curve. 

We then obtained dereddened colors $(V-I)_0$ and extinction-corrected mean magnitudes $I_0$ of the selected objects using reddening maps by \citet{skowron2021}. For the LMC fields, we removed stars located in the region of the CMD delimited by $(V-I)_0 \leq 0.5$ and $I_0 \leq 19.5$. This CMD region contains stars called ``blue-bumpers,'' exhibiting repeating outbursts that sometimes can be mistaken with microlensing events \citep[see,][]{alcock1997c,mroz2024a}. For the SMC fields, the brightness limit was slightly fainter, $I_0 \leq 20.0$, as the SMC distance modulus is 0.5\,mag larger than that of the LMC.

When analyzing selected light curves, we noticed that some apparent bumps were present during the same night in the light curves of many stars located in the same field or on the same CCD detector. These were spurious nonastrophysical signals produced by artifacts that were present in the reduced CCD images. They may have been caused by reflections of the light within the telescope optics or flawed calibration images (flats). We thus discarded stars that showed bumps during a few ``bad'' nights. After these steps, we were left with 308 light curves.

\subsection{Cut 3}

In the last step, we fitted microlensing models to the light curves of the remaining objects and selected brightenings that were consistent with microlensing. We considered three types of models. In the standard, point-source point-lens (PSPL) model, the magnification is given by the following formula:
\begin{equation}
A_{\rm PSPL}(t) = \frac{u(t)^2+2}{u(t)\sqrt{u(t)^2+4}},
\end{equation}
where $u(t)=\sqrt{\left(\left(t-t_0\right)/t_{\rm E}\right)^2+u_0^2}$ is the source--lens separation expressed in the angular Einstein radii units, $t_0$ is the moment of the closest lens--source approach, $t_{\rm E}$ is the Einstein radius crossing timescale, and $u_0$ is the impact parameter. The observed flux is given by 
\begin{equation}
F(t) = F_0 \left[1 + f_{\rm s}(A_{\rm PSPL}(t)-1)\right],
\end{equation}
where $F_0$ is the baseline flux and $f_{\rm s}$ is the dimensionless blending parameter. We considered two cases. If we fix $f_{\rm s}=1$ (four-parameter model), we assume that the whole observed flux comes from the source star. If we allow $f_{\rm s} \neq 1$ (five-parameter model), then we take into account the possibility of blended light that is not magnified during the event and may come from the lens itself or unrelated stars within the seeing disk of the event.

In the extended-source point-lens (ESPL) model, which takes into account the finite-source effects \citep{gould1994,mao1994,nemiroff1994}, we allowed the possibility that the angular radius of the source star $\theta_*$ is not negligible compared to the angular Einstein radius. Apart from the three standard parameters ($t_0$, $t_{\rm E}$, and $u_0$), the magnification depends on the normalized source radius $\rho\equiv\theta_*/\thetaE$ and can be calculated by integrating $A_{\rm PSPL}$ over the source area:
\begin{equation}
A_{\rm ESPL}(u,\rho) = \frac{\iint A_{\rm PSPL} dS}{\iint dS}.
\end{equation}
In practice, this integral can be tabulated \citep[e.g.,][]{gould1994,bozza2018}. Because there exists a mathematical degeneracy between the parameters of the ESPL model and the blending parameter \citep{mroz2020b,johnson2022}, we only considered ESPL models with fixed $f_{\rm s}=1$. We also neglected the limb darkening.

We found the best-fit parameters by minimizing the function
\begin{equation}
\chi^2 = \sum_{i}\frac{(F_i-F_0 \left[1 + f_{\rm s}(A_{\rm PSPL/ESPL}(t)-1)\right])^2}{\sigma^2_i},
\end{equation}
using the GNU Scientific Library\footnote{\url{https://www.gnu.org/software/gsl/}} implementation of the Levenberg--Marquardt algorithm \citep{levenberg1944,marquardt1963}. The algorithm returned not only the best-fit values but also their covariance matrix. We fitted three types of models: four- and five-parameter PSPL and five-parameter ESPL. However, we adopted only one of them as the ``fiducial'' model, using the following algorithm. 

If the finite-source effects were present in the light curve and the ESPL fit resulted in a significantly smaller $\chi^2$ than the PSPL fits, we adopted the ESPL fit as fiducial. Specifically, we required that \mbox{$\Delta\chi^2 \equiv \chi^2_{\rm PSPL} - \chi^2_{\rm ESPL} \geq 16$} (we checked that changing this parameter within a reasonable range did not affect our results). If $\Delta\chi^2<16$ and the five-parameter PSPL fit did not converge or it converged to an unphysical value of the blending parameter ($f_{\rm s}-\sigma(f_{\rm s}) \geq 1$), then we used the four-parameter PSPL fit instead. Otherwise, we adopted the five-parameter PSPL fit.

Once we identified the appropriate model, we calculated several goodness-of-the-fit statistics. We evaluated $\chi^2$ for the entire data set. In addition to that, we calculated $\chi^2_{\tE}$, $\chi^2_{2\tE}$, and $\chi^2_{\rm bump}$, the value of the $\chi^2$ statistics evaluated using only data points in the time ranges $|t-t_0|<\tE$, $|t-t_0|<2\tE$, or within the bump, respectively. We required that all four statistics fulfill the condition $\chi^2/\mathrm{d.o.f.}\leq2$ (where d.o.f. is the number of degrees of freedom), meaning that the model describes the data well. In addition to that, we required that the event timescale is reasonably well measured ($\sigma(\tE)/\tE\leq 1$), the source is brighter than $I_{\rm s} = 22$, and that the impact parameter $u_0 \leq \max\{1,2\rho\}$.

All criteria were met by only two objects, which we describe in more detail below. Their properties are summarized in Table~\ref{tab:events}. We treat one of them as a microlensing event candidate and call it OGLE-LMC-20. Furthermore, we argue that the second object is most likely a stellar flare. In addition to that, we visually inspected the light curves of all 2538 objects that met the ``cut 1'' criteria. However, we did not find any additional candidates for microlensing events.

\begin{figure}
\includegraphics[width=.5\textwidth]{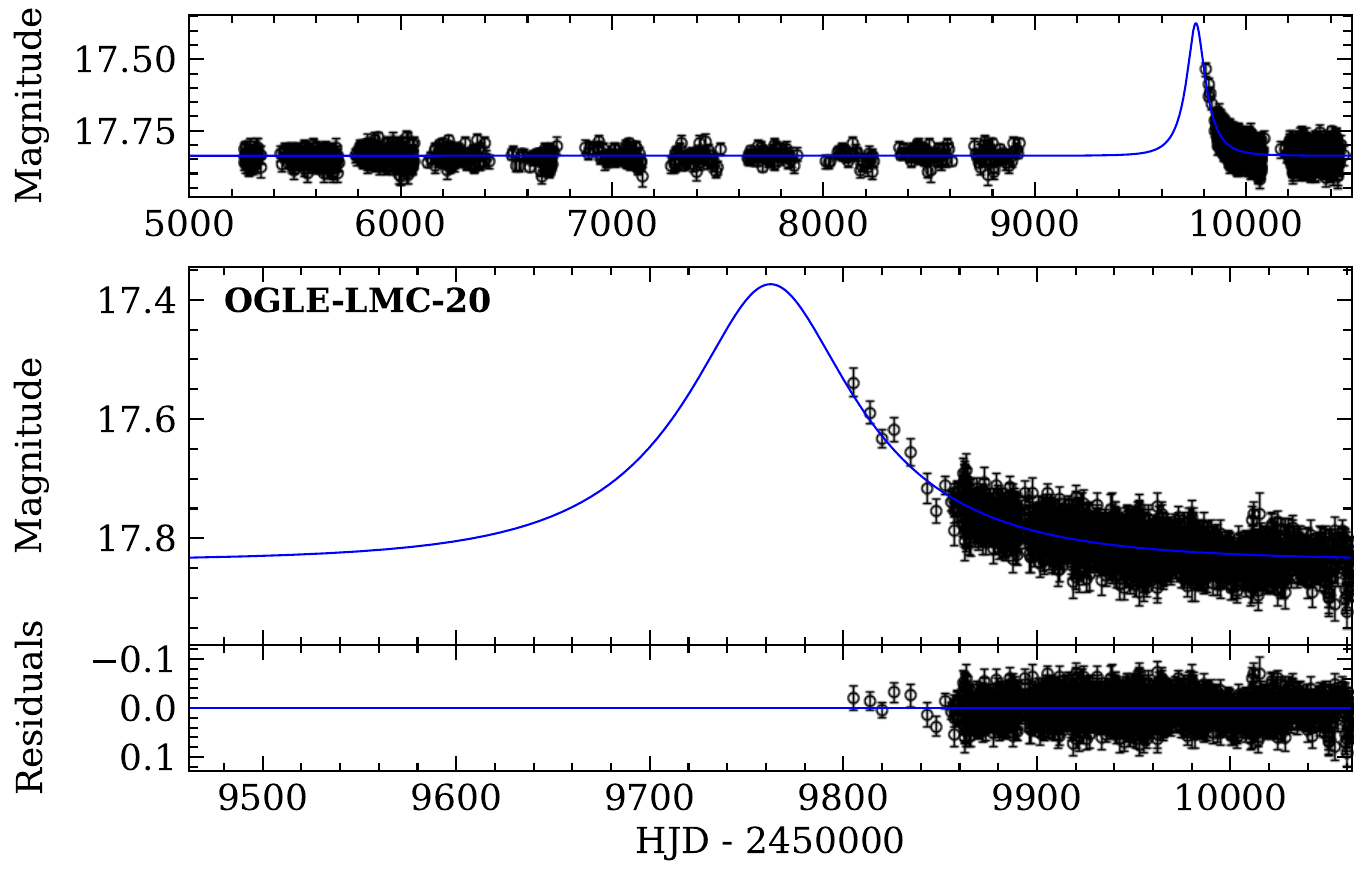}
\includegraphics[width=.5\textwidth]{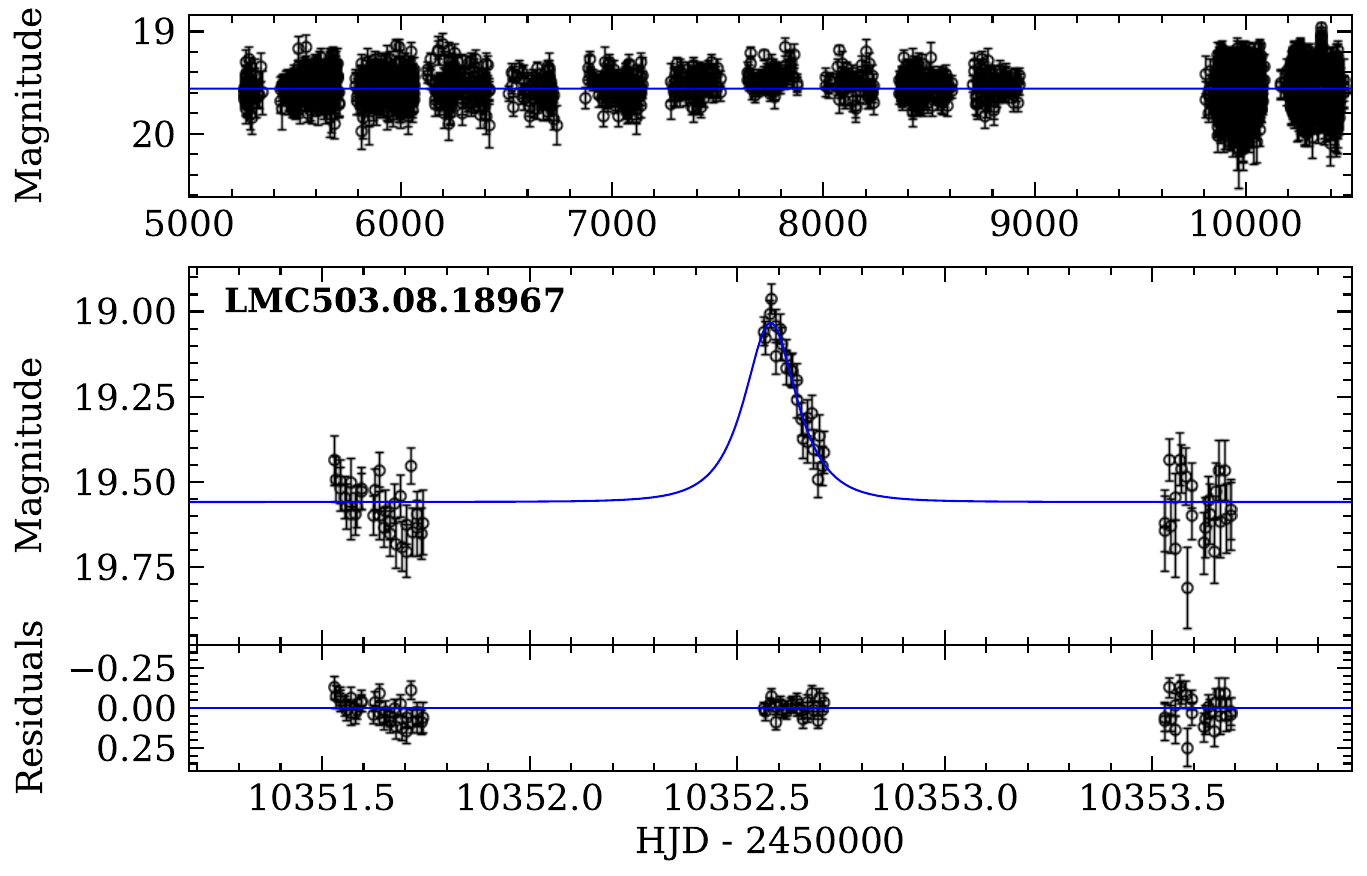}
\caption{Light curves of objects that passed all selection criteria. The solid blue line shows the best-fit PSPL microlensing model.}
\label{fig:lc}
\end{figure}

\begin{figure*}
\centering
\includegraphics[width=.7\textwidth]{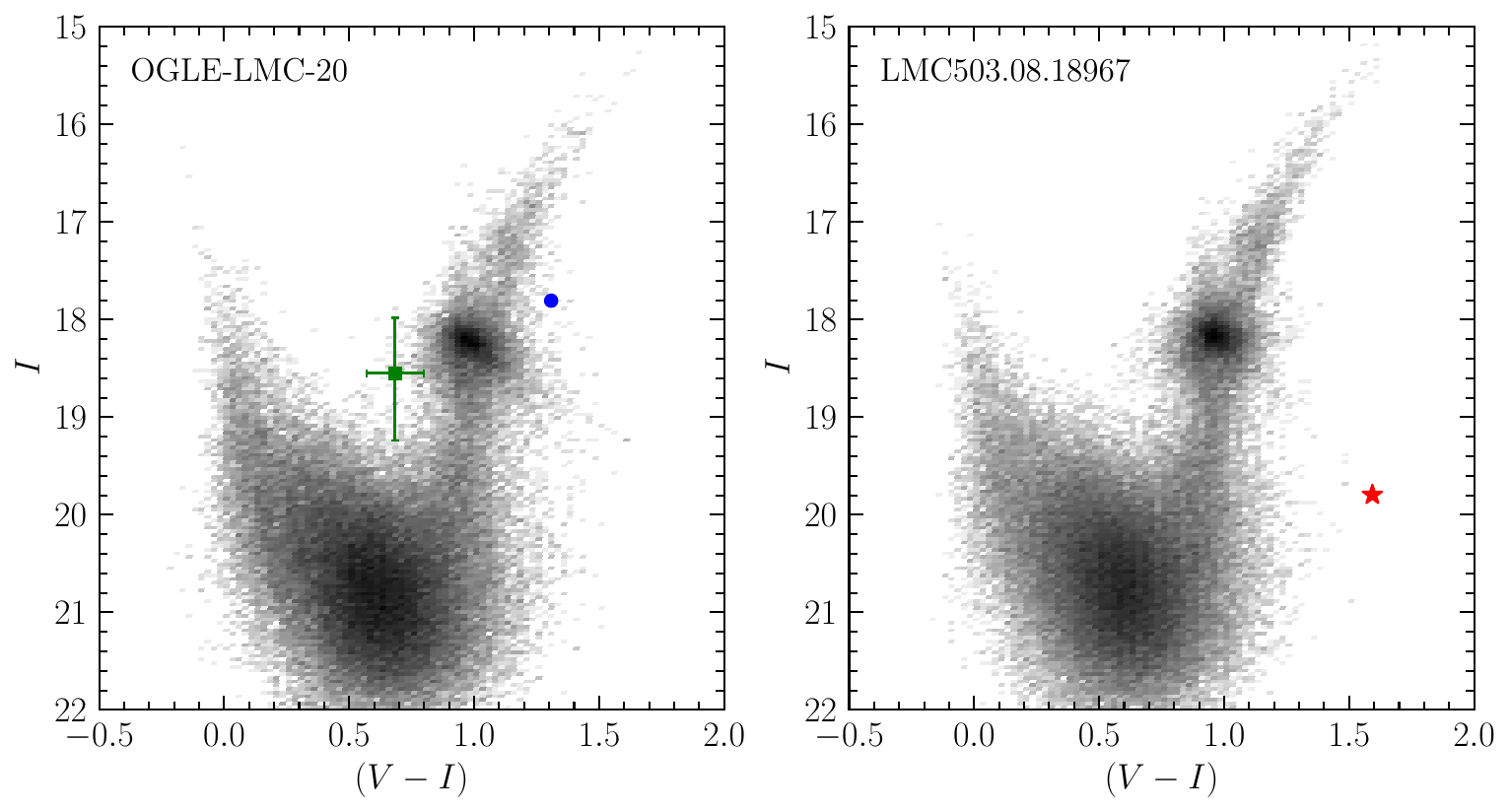}
\caption{Color--magnitude diagrams of the fields LMC503.15 (left panel) and LMC503.08 (right). The location of OGLE-LMC-20 (LMC503.08.18967) in the baseline is marked by a blue circle (red asterisk). The green square marks the position of the source star of OGLE-LMC-20.}
\label{fig:cmd}
\end{figure*}

\begin{deluxetable}{lrr}
\tablecaption{Detected Event Candidates\label{tab:events}}
\tablehead{
\colhead{} & \colhead{OGLE-LMC-20} & \colhead{LMC503.08.18967}}
\startdata
R.A. (J2000)        & \ra{05}{12}{57}{62}    & \ra{05}{26}{46}{31}  \\ 
Decl. (J2000)       & \dec{-69}{25}{21}{0}   & \dec{-69}{27}{38}{4} \\
OGLE-IV ID          & LMC503.15.81584        & LMC503.08.18967 \\
                    &                        & LMC516.32.58764 \\
$I_{\rm base}$      & $17.805 \pm 0.010$ & $19.795 \pm 0.016$ \\
$(V-I)_{\rm base}$  & $1.308 \pm 0.018$ & $1.593 \pm 0.048$\\
$t_0$               & $9752^{+15}_{-14}$  & $10352.5791^{+0.0058}_{-0.0085}$ \\
$\tE$ (day)         & $90^{+13}_{-10}$       & $0.104^{+0.044}_{-0.021}$ \\
$u_0$               & $0.47^{+0.20}_{-0.15}$ & $0.62^{+0.21}_{-0.26}$ \\
$I_{\rm s}$         & $18.55^{+0.69}_{-0.57}$& $19.86^{+0.90}_{-0.56}$ \\
$f_{\rm s}$         & $0.52^{+0.36}_{-0.24}$ & $0.76^{+0.52}_{-0.43}$ \\
\enddata
\tablecomments{$I_{\rm s}$ denotes the brightness of the source star in magnitudes.}
\end{deluxetable}

\section{Candidate Microlensing Events}

\subsection{OGLE-LMC-20}

The light curve of OGLE-LMC-20 is presented in the upper panels of Figure~\ref{fig:lc}. The brightening started between 2020 March and 2022 August, during a break in OGLE-IV operations. We treat this object as a microlensing event candidate because only the declining part of the light curve is covered by observations. The available data can be well modeled by a PSPL microlensing model with the Einstein timescale $\tE=90^{+13}_{-10}$\,day. This value is typical for the known microlensing events in the LMC direction \citep{mroz2024a}. This event occurred on a star at equatorial coordinates of (\ra{05}{12}{57}{62}, \dec{-69}{25}{21}{0}). Its mean brightness and color in the baseline are $I=17.805 \pm 0.010$ and $V-I=1.308 \pm 0.018$, respectively, which places the star close to the red giant branch in the CMD (that is shown in the left panel of Figure~\ref{fig:cmd}). However, the color of the source star, calculated using the model-independent regression, is smaller $(V-I)_{\rm s}=0.684 \pm 0.115$. The position of the source in the CMD is marked by a green square in the left panel of Figure~\ref{fig:cmd}.

\subsection{LMC503.08.18967}

This star exhibited a short-duration brightening during the night of 2024 February 11/12. Its light curve is presented in the lower panels of Figure~\ref{fig:lc}. The object did not undergo any additional brightenings nor exhibit periodic variability in the archival OGLE data (dated back to 1997). The star has equatorial coordinates of (\ra{05}{26}{46}{31}, \dec{-69}{27}{38}{4}) and is located in two partially overlapping fields: LMC503.08 and LMC516.32. The brightening was detected independently in the two light curves available for this object.

The light curve can be modeled by a PSPL model with a very short Einstein timescale $t_{\rm E}= 0.104^{+0.044}_{-0.021}$\,day. However, its rising part is unavailable, rendering it impossible to check for the symmetry of the bump and raising the possibility that the object was not the microlensing event at all. In fact, two lines of evidence indicate that the observed brightening was a stellar flare.

First, the observed color of the star ($V-I=1.593 \pm 0.048$) is much redder than the colors of main-sequence stars and giants from the LMC at similar apparent magnitudes (see the CMD in the right panel of Figure~\ref{fig:cmd}). This color index is consistent with colors of late K-type main-sequence stars with absolute magnitudes in the $I$ band of $M_I = 6.67^{+0.07}_{-0.15}$ \citep{pecaut2013}. If located in the LMC, such a star would be well below the OGLE limiting magnitude. Its apparent brightness indicates a much closer distance of $4.2^{+0.3}_{-0.1}$\,kpc, placing it in the Milky Way stellar halo or thick disk. Late K-type dwarfs are also known to be chromospherically active, which is consistent with the stellar flare interpretation.

The star was also detected by the VISTA survey of the Magellanic Clouds (\citealt{cioni2011,rubele2018}). Its near-infrared magnitudes ($Y=19.093 \pm 0.043$, $J=18.622 \pm 0.049$, $K_{\rm s}=17.731 \pm 0.094$) are also consistent with those expected from late K or early M dwarfs \citep{pecaut2013}.

\begin{figure}
\includegraphics[width=.5\textwidth]{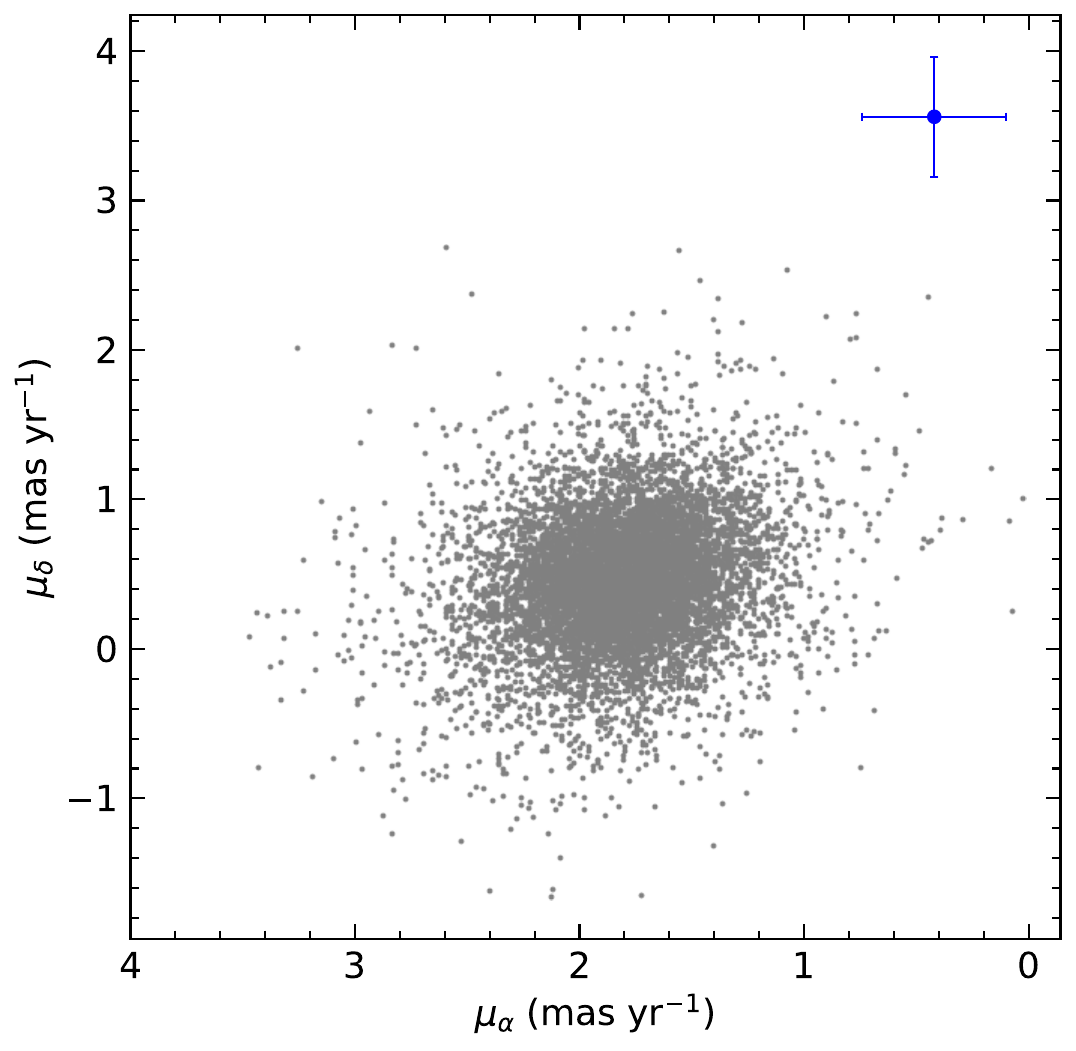}
\caption{Gray dots mark the proper motions of red clump stars in the field LMC503.08. The proper motion of LMC503.08.18967 is marked with a blue circle.}
\label{fig:pm}
\end{figure}

Second, the proper motion of the star is inconsistent with the LMC proper motion, as shown in Figure~\ref{fig:pm}. This object is not included in the \textit{Gaia} Data Release 3 \citep{gaia2016,gaia_edr3}. However, we were able to measure its proper motion using the archival OGLE data as part of the OGLE-Uranus project (Udalski et al., in preparation). The proper motion of the star is $(\mu_{\alpha},\mu_{\delta})=(0.42 \pm 0.32, 3.56 \pm 0.40)$ mas\,yr$^{-1}$. On the other hand, the LMC proper motion, defined here as the mean proper motion of nearby red clump stars,\footnote{We used stars located in the CMD region defined by $|I-I_{\rm RC}| \leq 0.6$ and $|(V-I)-(V-I)_{\rm RC}| \leq 0.2$, where $I_{\rm RC}$ and $(V-I)_{\rm RC}$ are the apparent magnitude and color of the red clump centroid.} is $(\mu_{\alpha},\mu_{\delta})_{\rm RC} = (1.82 \pm 0.35, 0.45 \pm 0.44)$ mas\,yr$^{-1}$. (The latter uncertainties quantify the rms scatter around the mean.) The proper motion of the star in the east and north direction differs by $3.0\sigma$ and $5.2\sigma$, respectively, from the LMC proper motion. The $p$-value of the hypothesis that the proper motion is consistent with that of the LMC is just $5\times 10^{-10}$, so this hypothesis can be ruled out at $6.2\sigma$ confidence level.


\section{Event Detection Efficiency}
\label{sec:efficiency}

We carried out extensive light-curve level simulations to calculate the event detection efficiency in the high-cadence survey. Their methodology is similar to that used in our previous works \citep{mroz2019b,mroz2024a}. It involves injecting the microlensing signal on top of the light curves of stars randomly drawn from the database, thereby preserving sampling, variability, and noise in the original data that would otherwise be difficult to simulate. We then check if the simulated light curves pass all the selection criteria discussed in Section~\ref{sec:search} (and summarized in Table~\ref{tab:cuts}).

Because we need to take into account finite-source effects, the detection efficiency $\varepsilon(\tE,\rho)$ is calculated as a function of both the Einstein timescale $\tE$ and the normalized source radius $\rho$. Finite-source effects affect the detectability of microlensing events in at least three ways. First, if $\rho \gg 1$, the duration of the event is set by the time needed for the lens to cross the source's surface, making it easier to detect small $\thetaE$ (short $\tE$) events. At the same time, the amplitude of the event (in magnitudes) decreases with increasing source radius as $2.171/\rho^2$. If $\rho > 6.6$, the amplitude of the event falls below 0.05\,mag, which was a threshold adopted for selecting candidate events (Table~\ref{tab:cuts}). Finally, the cross section to microlensing is no longer set by the angular Einstein radius but rather the angular size of the source star $\theta_*$.

The simulations were carried out for nine values of the normalized source radius ($\log\rho \in \{-3, -2, -1, 0, 0.2, 0.4, 0.6, 0.8, 1\}$). The peak time $t_0$ was drawn from a uniform distribution from the range $2,459,750 \leq t_0 \leq 2,460,450$ (i.e., between 2022 June 19.5 and 2024 May 19.5). The impact parameter was drawn from a range $0 \leq u_0 \leq u_{\rm max}$, where $u_{\rm max} = \max\{1, 2\rho\}$ to take into account all possible source-crossing events, and timescales were drawn from a log-uniform distribution from the range $10^{-2}$ to $10^2$\,days. The dimensionless blending parameter was drawn from empirical distributions created by matching stars detected in OGLE reference frames and high-resolution Hubble Space Telescope images (see, \citealt{mroz2024a}).

We simulated 160,000 events per field for each value of $\rho$ and each field. Because we noticed that the detection efficiency was rapidly dropping when $\rho > 2$, we ran additional simulations for $\log\rho=0.4$ (320,000 events per field), $\log\rho=0.6$ (480,000 events per field), and $\log\rho=0.8$ (640,000 events per field). The detection efficiency was calculated as the ratio of the number of events detected to the number of simulated events with $u_0 \leq 1$. 

\begin{figure}
\includegraphics[width=.5\textwidth]{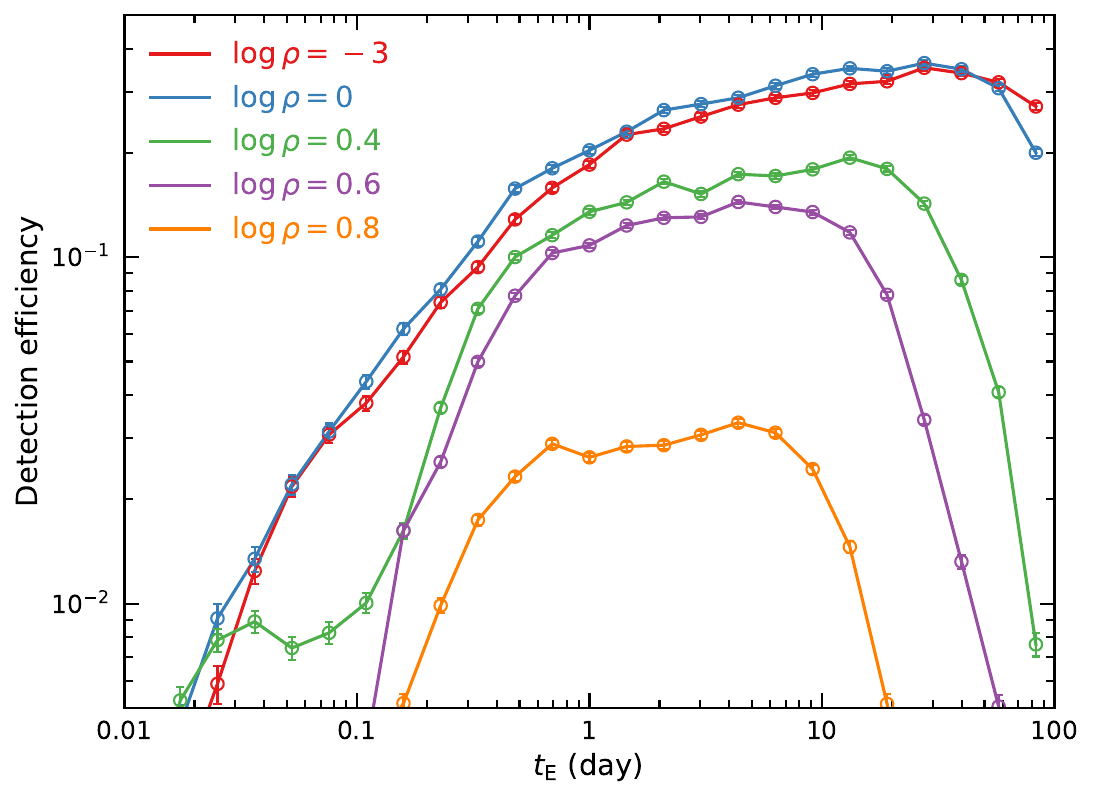}\\
\includegraphics[width=.5\textwidth]{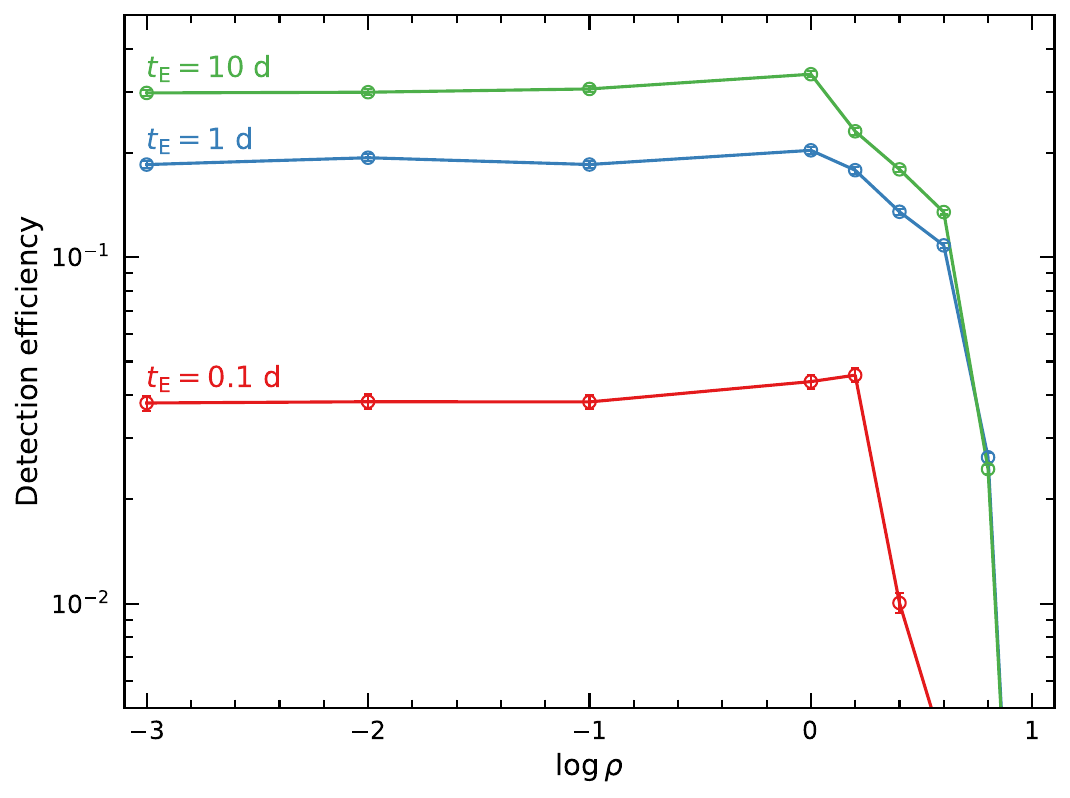}
\caption{Event detection efficiency in the field LMC509 as a function of Einstein timescale (upper panel) and normalized source radius (lower panel).}
\label{fig:eff}
\end{figure}

Example detection efficiency curves for the field LMC509 are shown in Figure~\ref{fig:eff}. If the finite-source effects are neglected, our survey has the highest sensitivity to events with timescales $\tE \approx 30$\,day (upper panel of Figure~\ref{fig:eff}). However, the event detection efficiency curve is flat, and the probability of finding events with $\tE \geq 1$\,day is larger than 50\% of the maximum detection efficiency. For shorter timescales, the probability of finding events decreases with decreasing event timescales. Including finite-source effects affects the detectability of events. The detection efficiency at $\rho \approx 1$ is elevated compared to the case $\rho \ll 1$, and it rapidly drops once $\rho \gtrsim 2\!-\!3$, when the finite-source effects tend to lower the amplitude of the event (as presented in the lower panel of Figure~\ref{fig:eff}).

\section{Limits on Planetary-mass PBHs}

During the presented high-cadence survey of the Magellanic Clouds, we detected only one candidate event (OGLE-LMC-20) with a relatively long timescale ($\tE=90^{+13}_{-10}$\,day). We will demonstrate below that this is well below the expected number of microlensing events from compact objects in dark matter. In fact, this sole event is consistent with the expected number of microlensing events from known stellar populations in the Milky Way disk and the LMC itself.

We calculate the limits on the abundance of planetary-mass PBHs in dark matter as a function of their mass $M$, assuming a monochromatic (delta-function) mass function. We denote the fraction of dark matter in the form of PBHs as $f=M_{\rm PBH}/M_{\rm DM}$, where $M_{\rm DM}$ is the total mass of dark matter halo and $M_{\rm PBH}$ is the total mass of PBHs within it. Our methodology closely follows that presented in \citet{mroz2024b}. 

We use the Milky Way and LMC disk and halo models to calculate the differential event rate $d^2\Gamma/d\tE d\rho$ and calculate the expected number of events:
\begin{equation}
N_{\rm exp} (f,M) = N_{\rm s}\Delta t \int \frac{d^2\Gamma}{d\tE d\rho}(\tE,\rho,f,M)\varepsilon(\tE,\rho)d\tE d\rho,
\end{equation}
where $N_{\rm s}$ is the number of source stars brighter than $I=22$ and $\Delta t=700$\,day is the duration of the survey. The number of source stars used in calculations is reported in Table~\ref{tab:fields} and taken from \citet{mroz2024a} for the LMC and Mróz et al.~(2025, in preparation) for the SMC. The integral is evaluated by generating an ensemble of $10^6$ events using the method described by \citet{clanton2014}. The finite-source effects are taken into account by assigning each simulated event a random source brightness $I_{\rm s}$ drawn from the luminosity function of a given field. Then, the angular radius of the source is calculated from
\begin{equation}
\theta_* = 
\begin{cases}
0.91\,{\mu}\text{as}\,10^{-0.2(I_{\rm s}-I_{\rm RC})} & \text{for LMC fields},\\
0.71\,{\mu}\text{as}\,10^{-0.2(I_{\rm s}-I_{\rm RC})} & \text{for SMC fields},
\end{cases}
\end{equation}
where $I_{\rm RC}$ is the mean brightness of red clump stars in this field and $\rho=\theta_*/\thetaE$. The angular radius of red clump stars is calculated using data from \citet{nataf2021}.

\begin{figure}
\includegraphics[width=.5\textwidth]{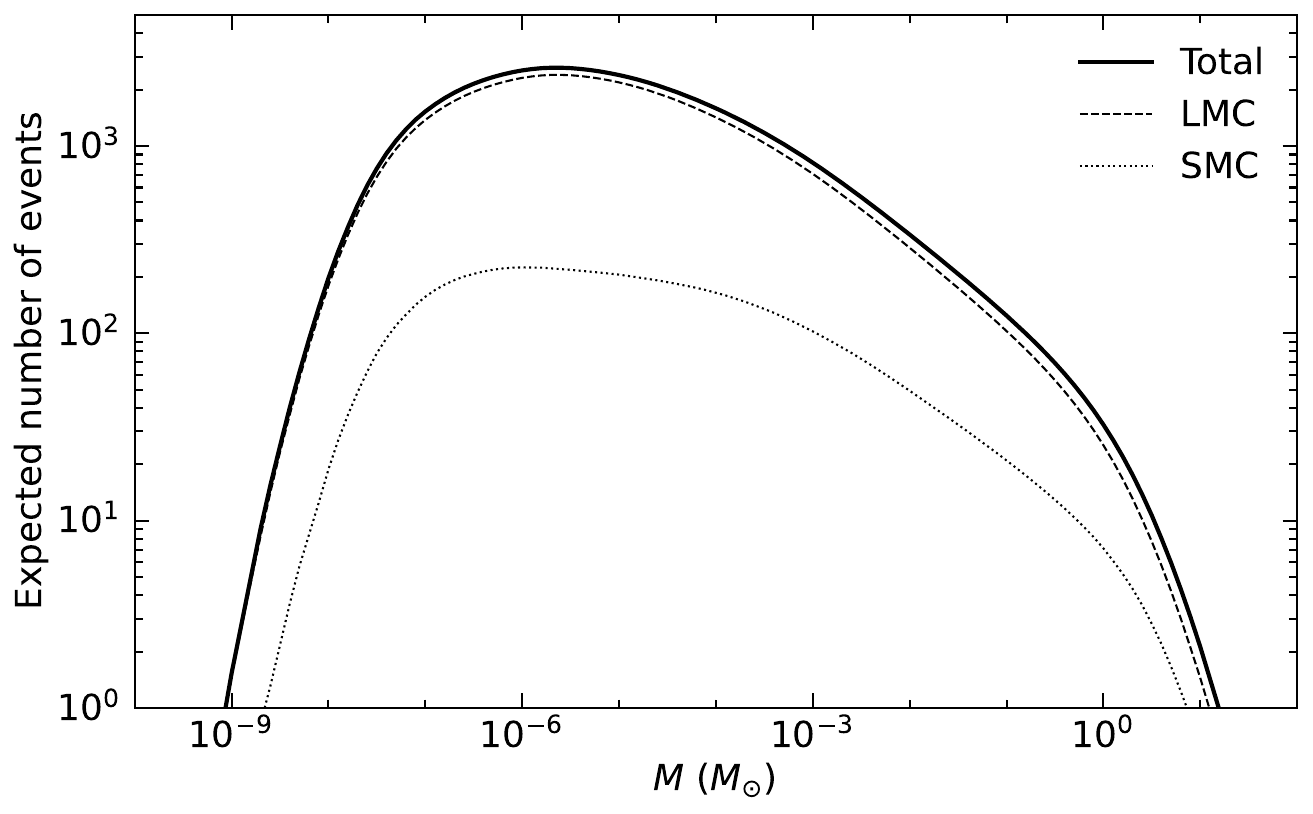}
\caption{Expected number of microlensing events in the OGLE High-cadence Magellanic Cloud Survey assuming that the entire dark matter is composed of PBHs or other compact objects of a given mass. The dashed and dotted lines mark the contributions from the sources in the LMC and SMC, respectively.}
\label{fig:nexp}
\end{figure}

The number of microlensing events expected to be detected in our survey, assuming that the entire dark matter is made of compact objects, is presented in Figure~\ref{fig:nexp}. We sum the contributions from the Milky Way halo (which is assumed to follow the \citet{cautun2020} contracted halo model with the total mass of $0.97\times 10^{12}\,M_{\odot}$ within 200\,kpc) and the LMC halo (which is modeled by a Hernquist profile with a total mass of $1.49 \times 10^{11}\,M_{\odot}$; \citealt{erkal2019}). These models are described in more detail in \citet{mroz2024b}. We neglect the contribution from the SMC dark matter halo because of its small mass ($\approx 6.5\times 10^9\,M_{\odot}$; \citealt{bekki2009}), relatively small number of sources in the SMC (four times smaller than in the LMC), and shorter duration of the SMC high-cadence survey (236 versus 435 nights).

Our survey has the highest sensitivity to PBHs of $2.5\times 10^{-6}\,M_{\odot}$. We should have detected 2615 microlensing events if the entire dark matter was composed of compact objects of that mass. The expected number of events falls off steeply for lower masses due to finite-source effects. Still, we should have found more than 100 short-timescale microlensing events if dark matter was made of PBHs in the planetary- and brown-dwarf-mass range (from $6.6\times 10^{-9}\,M_{\odot}$ to $0.16\,M_{\odot}$).

On the other hand, if we take into account only known stellar populations in the Milky Way disk, we should have detected from 0.4 to 0.9 microlensing events depending on the adopted disk model (\citealt{han_gould2003} or \citealt{cautun2020}, respectively). The expected number of self-lensing events in the LMC is only 0.1. Thus, the total number of expected events is 0.5--1.0, in good agreement with one event detected. The timescale of this event ($\tE=90^{+13}_{-10}$\,day) also matches that expected from the Milky Way and LMC stellar populations (see Extended Data Figure 6 in \citealt{mroz2024b}).

Assuming that OGLE-LMC-20 originates from the lens in the Milky Way disk or in the LMC, we calculate the posterior distribution for the frequency of PBHs in dark matter $f$ from the Bayes' formula:
\begin{equation}
P(f|M) \propto \mathcal{L}(f,M) P_0(f),
\end{equation}
where
\begin{equation}
\mathcal{L}(f,M) = \mathrm{e}^{-N_{\rm exp}(f,M)}
\end{equation}
is the likelihood function and $P_0(f)$ is a flat (uniform) prior on $f$. This calculation is repeated for 121 logarithmically spaced masses ranging from $10^{-10}\,M_{\odot}$ to $10^2\,M_{\odot}$ using the methodology developed by \citet{mroz2024b}.

The 95\% confidence upper limits on $f$ are presented by a thick purple line in Figure~\ref{fig:bounds}. The new limits are strongest for $M \approx 10^{-6}\,M_{\odot}$ for which they reach almost $f\approx 10^{-3}$. Compact objects in the mass range from $1.4\times 10^{-8}\,M_{\odot}$ (half of the Moon mass) to $0.013\,M_{\odot}$ (planet/brown dwarf boundary) can comprise at most 1\% of dark matter. At the low-mass end, we are fundamentally limited by the finite-source effects, which render it difficult to detect low-mass (low $\thetaE$) events. A possible solution to overcome this limitation would be to observe fainter (that is, smaller) source stars in the LMC or change the target to a more distant galaxy (for example, the Andromeda galaxy). Still, masses smaller than $\approx 10^{-10}\,M_{\odot}$ cannot be probed by optical microlensing because of the wave-optics effect, a situation when the Schwarzchild radius of a PBH becomes comparable to the optical wavelengths \citep{sugiyama2020}. 

When our current results are combined with those from the Subaru/HSC survey \citep{niikura2019} for low PBH masses, and those from the combined \mbox{OGLE-III} and \mbox{OGLE-IV} data set \citep{mroz2024b} at the high-mass end, we find that compact objects in the mass range from $3.5\times 10^{-10}\,M_{\odot}$ to $6.3\,M_{\odot}$ cannot make up more than 1\% of dark matter. Those in the $1.0\times 10^{-10}\,M_{\odot}$ to $860\,M_{\odot}$ range cannot comprise more than 10\% of dark matter. These limits span 13 orders of magnitude, virtually the entire range than can be realistically probed by microlensing.

Our results clearly rule out the claims by \citet{niikura2019b} that the short-timescale microlensing events detected in the direction of the Galactic bulge may be a signature of a sizable population of planetary-mass PBHs. These events are most likely caused by a population of free-floating or wide-orbit planets in the Milky Way.

The data presented in this paper are publicly available at \url{https://ftp.astrouw.edu.pl/ogle/ogle4/LMC_FFP_PBH/}.

\section*{Acknowledgements}

This research was funded in part by National Science Centre, Poland, grants OPUS 2021/41/B/ST9/00252 and SONATA 2023/51/D/ST9/00187 awarded to P.M.

\bibliographystyle{aasjournal}
\bibliography{pap}

\end{document}